%% file: dpf2013proc.tex
\documentclass[12pt]{article}
\usepackage{graphicx}
\usepackage{color}
\usepackage{array}
\usepackage{latexsym}
\usepackage[italic]{hepparticles}
\usepackage[italic]{heppennames}

\newcommand\pubnumber{DPF2013-118}
\newcommand\pubdate{\today}

\def\scipp{Santa Cruz Institute for Particle Physics,
University of California, Santa Cruz}

\def\Title#1{\begin{center} {\Large #1 } \end{center}}
\def\Author#1{\begin{center}{ \sc #1} \end{center}}
\def\Address#1{\begin{center}{ \it #1} \end{center}}

\newcommand\pubblock{\rightline{\begin{tabular}{l} \pubnumber\\
         \pubdate  \end{tabular}}}
\newenvironment{Abstract}{\begin{quotation}  }{\end{quotation}}
\newenvironment{Presented}{\begin{quotation} \begin{center} 
             PRESENTED AT\end{center}\bigskip 
      \begin{center}\begin{large}}{\end{large}\end{center} \end{quotation}}
\def\Acknowledgments{\bigskip  \bigskip \begin{center} \begin{large}
             \bf ACKNOWLEDGMENTS \end{large}\end{center}}

\input symbols.tex

\begin{document}
\begin{titlepage}
\pubblock

\vfill
\Title{Recent Results on Radiative and Electroweak Penguin Decays of 
\PB Mesons at \babar}

\vfill
\Author{ A.M. Eisner (for the \babar\ Collaboration)}
\Address{\scipp}
\vfill
\begin{Abstract}
  Radiative and electroweak decays of \PB mesong ($\PB\to\xsd\Pgg$ and
  $\PB\to\xsd\ellell$, respectively, where $X$ is a hadronic system and
  \Pqs or \Pqd labels the underlying quark process $\Pqb\to\Pqs(\Pqd)\Pgg$)
  provide good places to search for new physics (NP) beyond the
  standard model.  Recent \babar\ measurements are
  reported: inclusive $\BR(\PB\to\xs\Pgg)$, direct \CP asymmetry 
  for inclusive $\xs\Pgg$ and $\xsplusd\Pgg$, and 
  searches for the rare decays $\PB\to\Pgp(\Pgh)\ellell$.
  No evidence for NP was found, but several constraints
  are discussed.
\end{Abstract}
\vfill
\begin{Presented}
DPF 2013\\
The Meeting of the American Physical Society\\
Division of Particles and Fields\\
Santa Cruz, California, August 13--17, 2013\\
\end{Presented}
\vfill
\end{titlepage}
\def\thefootnote{\fnsymbol{footnote}}
\setcounter{footnote}{0}

\section{Introduction}

\begin{figure}[h]
 \begin{center}
  \includegraphics[width=0.35\textwidth,clip]{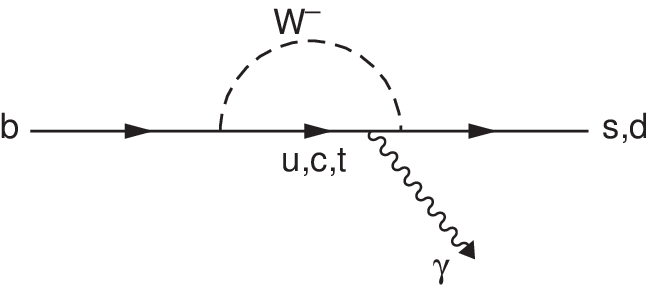} 
  \includegraphics[width=0.64\textwidth,clip]{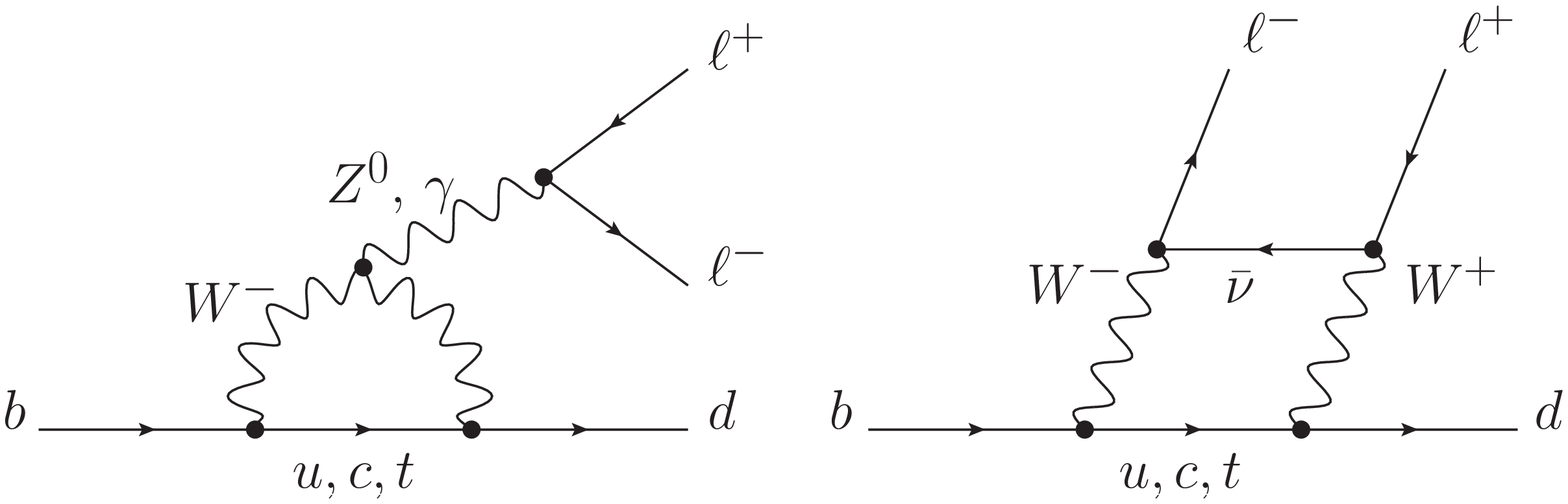}
 \end{center}
 \caption{Illustrative quark-level diagrams for radiative (left) and 
  electroweak (middle and right) decays of \PB mesons in the standard model.}
 \label{fig:diag}
\end{figure}

Unlike the dominant decays of \Pqb quarks, the flavor-changing neutral 
current processes $\Pqb\to\Pqs(\Pqd)\Pgg$ and $\Pqb\to\Pqs(\Pqd)\ellell$ 
(where \Plpm is \Pepm or \Pgmpm) do not occur at leading (tree) order, but
rather via the loop and box diagrams illustrated in Fig.~\ref{fig:diag}.
The corresponding hadron-level inclusive processes are written 
$\PB\to\xsd\Pgg$ and $\PB\to\xsd\ellell$.

These are rare decays, with small branching fractions (BF's). 
They are a good place to look for physics beyond the standard model (SM),
since new particles can appear virtually in the loops.
Extensive theoretical effort has led to low SM
uncertainties both for the predicted BF and direct \CP asymmetry \acp
in the inclusive decays, implying
good sensitivity to new physics (NP).  Predictions for
exclusive final states are less precise.

I present some recent measurements of
these processes by the \babar\ experiement at the PEP-II asymmetric
\epem collider (SLAC National Accelorator Laboratory):

\begin{enumerate} 
 \item Fully-inclusive measurement of $\PB\to\xs\Pgg$.~\cite{Lees:2012ym,
   Lees:2012ufa}
 \begin{itemize}
  \item $\BR(\PB\to\xs\Pgg)$ -- sensitive to NP.
  \item Direct \acp in $\PB\to\xsplusd\Pgg$ -- sensitive to NP.
  \item Photon energy spectrum in $\PB\to\xs\Pgg$ -- not sensitive to NP
   (rather, it reflects the motion of the quark inside the
   \PB, \ie, the shape function).
 \end{itemize}
 \item Direct \acp in $\PB\to\xs\Pgg$ via sum of exclusive modes --
               sensistive to NP \\
               (\babar\ preliminary results).
 \item Search for $\PB\to\xd\ellell$ decays in exclusive 
               modes -- SM predictions for \Pgp\ellell and \Pgh\ellell
               are $\mathcal{O}(1\ \mathrm{to}\ 4\times 10^{-8})$.
               ~\cite{Lees:2013lvs}
 \item Not included in this presentation is a search for
  $\PB\to\PK^{(*)}\Pgn\Pagn$ with hadronic recoil. 
  SM BFs are $\approx 4.5 (6.8) \times 10^{-6}$ for \PK(\PKst).
  The new \babar\ 90\% CL isospin-averaged limits are
  $\approx 32 (79) \times 10^{-6}\ $~\cite{Lees:2013kla}, but combine
  with earlier independent \babar\ results to give smaller values.
\end{enumerate}

\section{Some Theoretical Predictions}

The effective Hamiltonian for \PB decays
is written as a sum of operators $\mathcal{O}_i$ times Wilson coefficients, 
$C_i$.  For $\PB\to \xsd\Pgg$ in the SM, the relevant terms are those
involving $\mathcal{O}_1$, $\mathcal{O}_2$, $\mathcal{O}_7$, and 
$\mathcal{O}_8$, while for $\PB\to \xsd\ellell$ 
$\mathcal{O}_9$ and $\mathcal{O}_{10}$ also come into play.  In the SM
the coefficients are real; NP may introduce non-zero phases.

A computation based on renormalization-group techniques, involving thousands 
of diagrams and many physicists, has resulted in an SM prediction at
NNLO (next-to-next-leading-order) of
\begin{equation}
   \BR(\PB\to\xs\Pgg) = 
   (3.15 \pm 0.23)\times 10^{-4}\ (\eg > 1.6\gev), 
   \label{eq:NNLO}
\end{equation}
where \eg is the photon energy in the \PB rest 
frame -- see~\cite{Misiak:2006zs} and references therein.

Because the \Pqt quark dominates the loop in the radiative decay
(left plot of Fig.~\ref{fig:diag}),
\begin{equation}
  \BR(\bxdg)/\BR(\bxsg) \approx (\Vtd/\Vts)^2 = 0.044 \pm 0.003\, . 
  \label{eq:bfRatio}
\end{equation}

For direct \acp in $\PB\to\xsd\Pgg$, 
$$ \acp(\PB\to\xsd\Pgg) \equiv \frac
 {\Gamma(\PB\to\xsd\Pgg) - \Gamma(\PaB\to\xasd\Pgg)}
 {\Gamma(\PB\to\xsd\Pgg) + \Gamma(\PaB\to\xasd\Pgg)}\, , $$
older SM computations~\cite{Hurth:2003dk} predicted
$\acp(\PB\to\xsd\Pgg) = 0.0044^{+0.0024}_{-0.0014} \ 
(-0.102^{+0.033}_{-0.058})$.
If \xs and \xd are not separated, it was found that the combined
$$ \acp(\PB\to\xsplusd\Pgg) \equiv \frac
    {\Gamma(\PB\to\xs\Pgg + \PB\to\xd\Pgg)
    - \Gamma(\PaB\to\xas\Pgg + \PaB\to\xad\Pgg)}
     {\Gamma(\PB\to\xs\Pgg + \PB\to\xd\Pgg)
    + \Gamma(\PaB\to\xas\Pgg + \PaB\to\xad\Pgg)} = 0 $$
to a precision of $10^{-6}$, providing a very sensitive test for NP.

It has more recently been found~\cite{Benzke:2010tq} that
previously-ignored long-distance (``resolved-photon'') effects
increase the uncertainty on the separate $\acp(\PB\to\xsd\Pgg)$
predictions in the SM.  In particular, 
\begin{equation}
  -0.006 < \acp(\PB\to\xs\Pgg) < 0.028\ \mathrm{(SM\ prediction)}. 
  \label{eq:xsACP}
\end{equation}
However, the additional effects cancel for a newly-proposed
test for NP:
\begin{equation}
   \Delta\acp(\xs)\equiv \acp(\PBm\to\xs^-\Pgg) - \acp(\PB\to\xs^0\Pgg)
   = 4\pi^2 \alpha_s (\tilde{\Lambda}_{78}/m_{\Pqb}) 
   \mathrm{Im}(C_8/C_7) \, 
   \label{eq:DeltaACP}
\end{equation}
Because the Wilson coefficients are real in the SM, a non-zero result
implies NP; however, 
the hadronic factor $\Lambda_{78}$ can be anywhere between
17 and 190\mev.  In addition, the precise predictgion
of $\acp(\PB\to\xsplusd\Pgg) = 0$ is preserved in the new calculation.

\section{Fully-inclusive $\PB\to\xs\Pgg$}

\begin{figure}[htb]
 \begin{center}
  \includegraphics[width=0.6\textwidth]{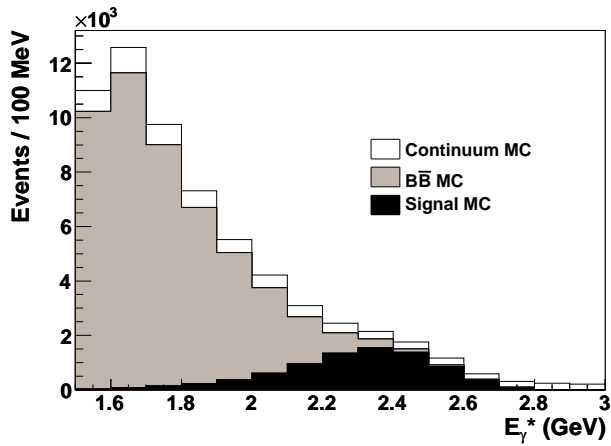}
 \end{center}
 \vspace*{-0.2in}
 \caption{Monte Carlo estimates (based on GEANT4, EVTGEN) of event spectra
  (scaled to data luminosity) after event selection, \babar\ 
  fully-inclusive $\PB\to\xs\Pgg$.}
 \label{fig:aftersel}
\end{figure}

For a fully-inclusive analysis, inclusivity is ensured by making no
requirements on the signal \PB meson other than on the
high-energy photon.  For the recent \babar\ 
analysis~\cite{Lees:2012ym, Lees:2012ufa}, the reconstructed photon 
has CM (\Y4S)-frame energy $\egcms >= 1.53\gev$.
For a given event, the \PB rest frame is not known;
\egcms differs from \eg by Doppler smearing (the motion of
the \PB in the CM frame) and calorimeter energy resolution.  
Backgrounds come from continuum events ($\epem\to \Pq\Paq\ (\Pq\neq\Pqb)\ 
\mathrm{or}\ \tautau$) and other \BB events (with the photon arising
from the decay of an intermediate particle or via misidentification).
Since many background photons come from $\Pgpz(\Pgh)\to\Pgg\Pgg$,
an event is vetoed if the partner \Pgg is found with appropriate
$m_{\Pgg\Pgg}$.  Continuum is additionally suppressed using full-event
topology, and by requiring a high-momentum lepton (\Pe or \Pgm) tag.
Signal and other \BB events may be tagged if the non-signal \PB decays
semileptonically; such a lepton is far less likely for
continuum.  As a bonus, the lepton provides a \CP tag.
Event spectra after all selection criteria are shown in
Fig.~\ref{fig:aftersel}.

The data consist of 347\invfb collected at the \Y4S resonance, and
36\invfb collected 40\mev below the \Y4S.  The continuum background is
subtracted by scaling the off-resonance data to the on-resonance sample;
this dominates the statistical uncertainty of the results.
The \BB background is subtracted using data-corrected Monte Carlo (MC)
computations.  Most of the dozen sources of \BB background (see 
Table~I in~\cite{Lees:2012ufa} for details) are corrected using 
dedicated studies of data \vs MC control samples.  The \BB background
is the dominant source of systematic uncertainty.  Because of large
\BB backgrounds, we obtain no useful signal result below 1.8\gev.
The signal region above 1.8\gev was kept ``blind'' during the
determination of all selection criteria and \BB corrections; the
range 1.53 to 1.8\gev served as a control region.  Lastly, for the
direct-\acp measurement, events were counted by lepton charge
for $\egcms > 2.1\gev$.

\begin{figure}[!b]
 \begin{minipage}[c]{0.5\textwidth}
  \begin{center}
   \includegraphics[width=0.99\textwidth]{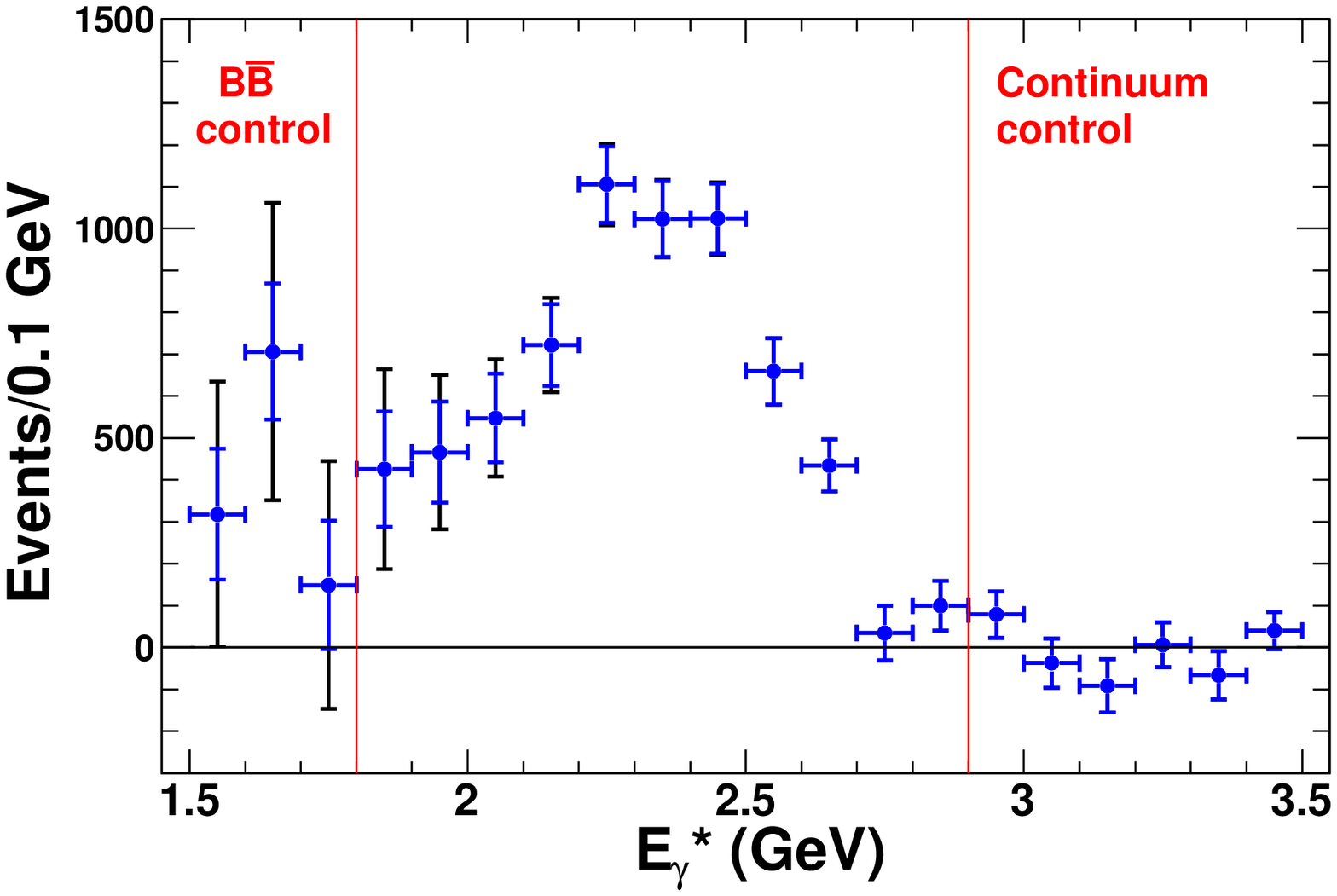} 
   \vspace*{-0.2in}
   \parbox{0.9\textwidth}{\caption{\babar\ fully-inclusvie photon spectrum 
    after background subtraactions.  Inner errors are statistical only,
    outer include systematics (which are highly correlated between bins)}
   \label{fig:rawsignal}}
  \end{center}
 \end{minipage}
 \begin{minipage}[c]{0.5\textwidth}
  \begin{center}
   \includegraphics[width=0.99\textwidth]{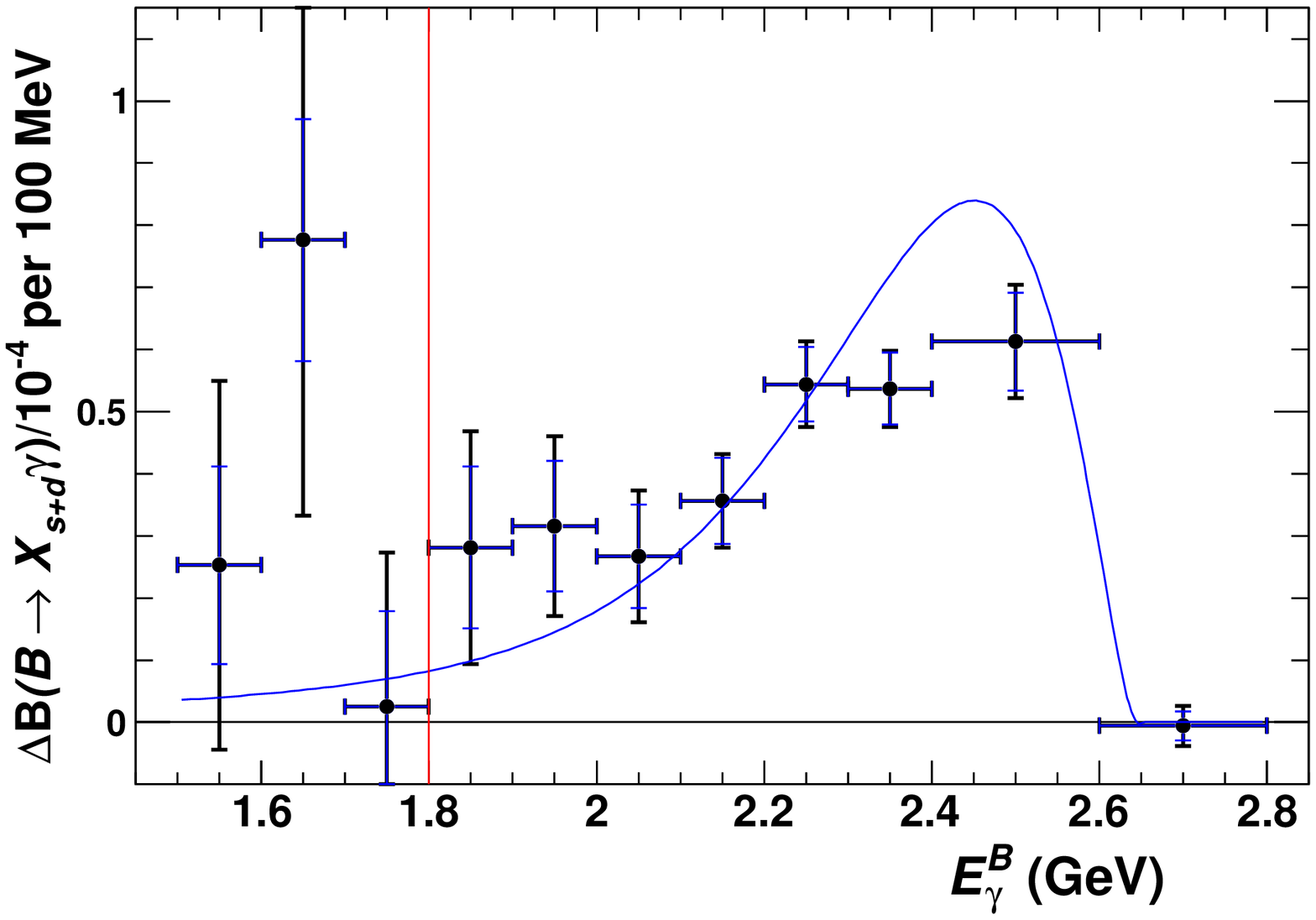}
   \vspace*{-0.2in}
   \parbox{0.9\textwidth}{\caption{\babar\ fully-inclusive photon spectrum
    unfolded to true \eg.  Errors are as in 
    Fig.~\ref{fig:rawsignal}.  The overlaid curve is based on
    HFAG's world-average HQET parameters (see text).}
   \label{fig:unfolded}}
  \end{center}
 \end{minipage}
\end{figure}

Figure~\ref{fig:rawsignal} shows the signal after background
subtaction.  The $\xs\Pgg$ BF for a given energy threshold is
found by correcting the signal yield in that region for efficiency, 
making a small adjustment from \egcms to \eg, including the
additional systematics, and scaling by 0.958
(Eq.~\ref{eq:bfRatio})
to account for $\xd\Pgg$.  For the lowest threshold:
\begin{equation}
 \BR(\PB\to\xs\Pgg) = (3.21 \pm 0.15 \pm 0.29 \pm 0.08) \times 10^{-4} \ 
 (\eg > 1.8\gev)
 \label{eq:xsBF18}
\end{equation}
The errors are statistical, systematic, and model-dependence
(``model'' for the true \eg spectrum), respectively.  Results are also
reported for thresholds of 1.9 and 2.0\gev.

We have additionally unfolded the measured \egcms spectrum to a spectrum
in true \eg, using a method adapted from B.~Malaescu~\cite{Malaescu:2009dm}.
The result (not directly related to NP) is shown in
Fig.~\ref{fig:unfolded}.  The overlaid curve shows the spectral
shape given by a Heavy Quark Effective Theory (HQET) computation in the
``kinetic scheme''~\cite{Benson:2004sg}, using world-average HQET
parameters determined by the Heavy Flavor Averaging Group (HFAG) from
$\PB\to X_{\Pqc}\ell\Pgn$ and $\PB\to\xs\Pgg$ data.

\begin{figure}[!t]
 \begin{minipage}[c]{0.52\textwidth}
  \begin{center}
   \includegraphics[width=\textwidth]{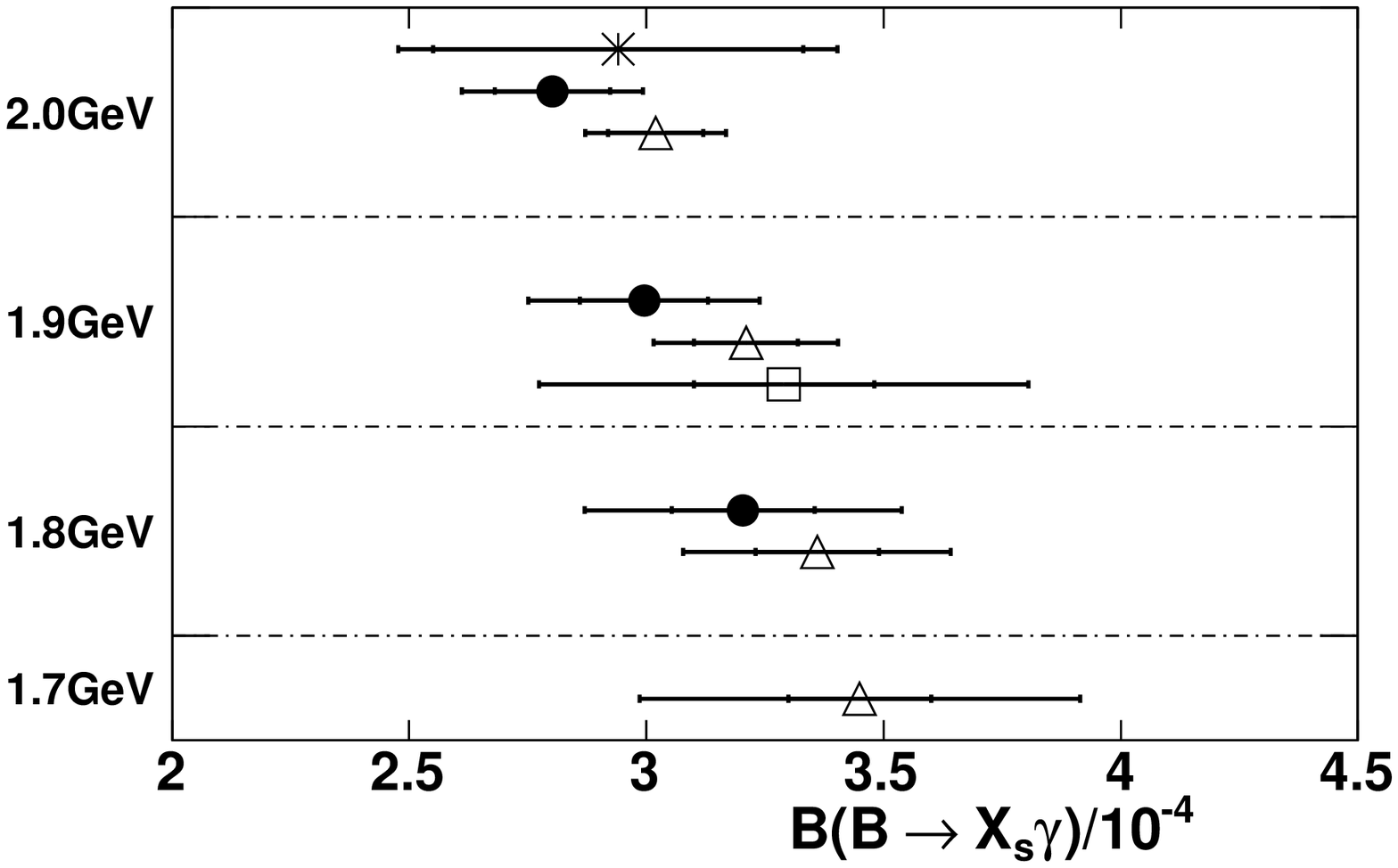}
   \parbox{0.9\textwidth}{\caption{Comparison of $\BR(\PB\to\xs\Pgg)$ 
    measurements for several
    \eg thresholds.  Shown are the current \babar\ measurement ({\large
    $\bullet$}), CLEO ({\large $\ast$}), Belle ({\small $\triangle$}) and 
    \babar\ sum-of-exclusive ({\small $\Box$}).}
   \label{fig:xsCompare}}
  \end{center}
 \end{minipage}
 \begin{minipage}[c]{0.48\textwidth}
  \begin{center}
   \includegraphics[width=\textwidth]{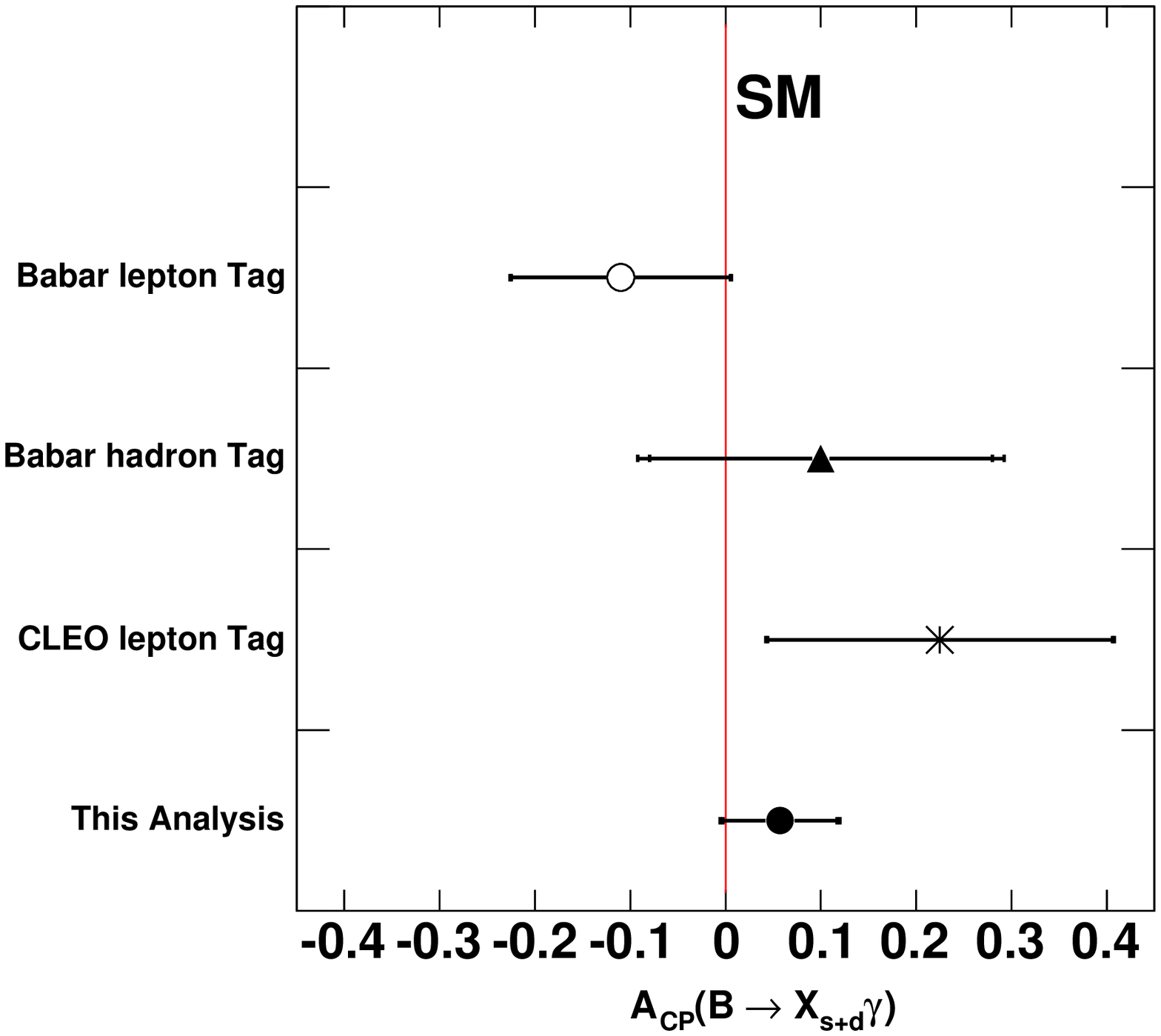}
   \parbox{0.9\textwidth}{\caption{Comparison of measurements of 
    $\acp(\PB\to\xsplusd\Pgg)$.
    ``\babar\ leptonn Tag'' is superceded.}
   \label{fig:acpCompare}}
  \end{center}
 \end{minipage}  
\end{figure}

Figure~\ref{fig:xsCompare} compares the BF results to earlier
experiments.  Measurements with different thresholds from a single
experiment are strongly correlated.  Uncertainties increase toward
lower thresholds due to increasing \BB backgrounds.

To compare the \babar\ result to the SM prediction of Eq.~\ref{eq:NNLO}, 
it must be extrapolated down to 1.6\gev.  The standard procedure
is to apply a factor provided by HFAG, based on
the same world-average HQET model described above.  While this
procedure (in particular its claimed uncertainty) does not seem
entirely satisfactory, we have applied it to obtain
$\BR(\PB\to\xs\Pgg) = (3.31\pm 0.35)\times 10^{-4}\ (\eg > 1.6\gev)$.
Such a result limits the amount of ``room'' for NP.  For example, in
a type-II Two-Higgs Doublet Model (THDM)~\cite{Misiak:2006zs},  
$m_\PHpm < 327\gevcc$ is excluded at 95\% C.L. for all values
of $\tan\beta$.  A recent THDM update~\cite{Hermann:2012fc} strengthens
the limit.

\section{\boldmath{$\acp(\PB\to\xsplusd\Pgg)$} from fully-inclusive data}

In contrast to the situation for the BF, \acp behaves very differently
for $\PB\to\xs\Pgg$ and $\PB\to\xd\Pgg$, so only the combined \acp
for \xsplusd can be measured by the fully-inclusive technique.  \PB
\vs \PaB is tagged by lepton charge.  After correcting for mistags
(primarily from \PB-\PaB mixing), we have measured
\begin{equation}
  \acp(\PB\to\xsplusd\Pgg) = 0.057 \pm 0.060\,(\mathrm{stat})
  \pm 0.018\,(\mathrm{syst})\ ,
  \label{eq:acpsplusd}
\end{equation}
consistent with the SM prediction of 0.  Figure~\ref{fig:acpCompare}
compares this result to earlier, less precise, measurements.

\section{\boldmath{$\acp(\PB\to\xs\Pgg)$} by Sum of Exclusive Decays}

With reconstructed (exclusive) final states, \xs is distinguished from
\xd by a \PKpm or \PKzS in the final state.  \CP charge can be assigned
for \PBp \vs \PBm decay, and for \PBz by \PKp \vs \PKm in the final state.
The \babar\ measurement, based on 420\invfb of data, uses the modes
listed in Table~\ref{tab:CPmodes}.

\input CPmodes.tex

\begin{figure}[!t]
 \begin{center}
  \includegraphics[width=0.6\textwidth]{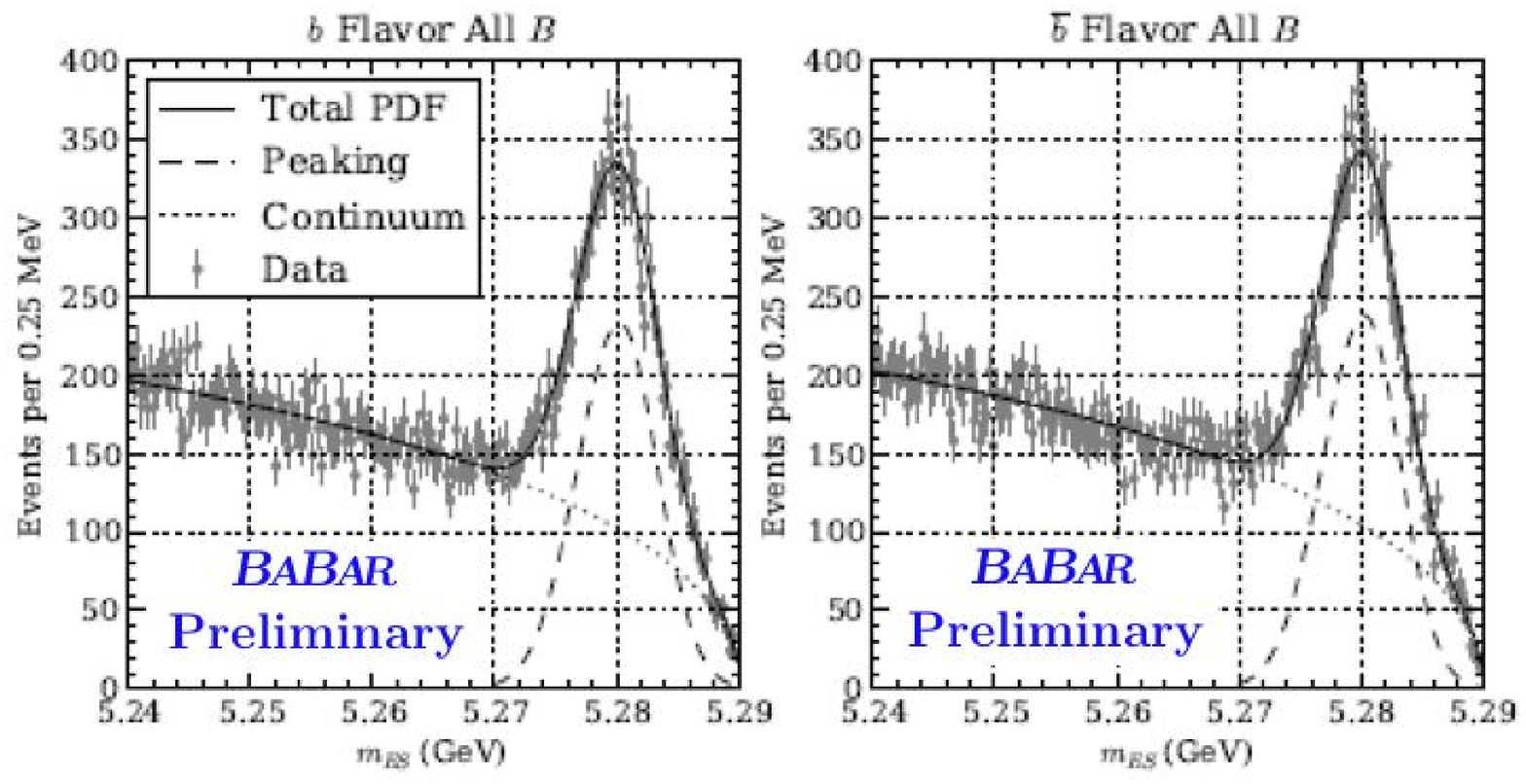}
  \caption{\mes spectra and fits for \PB (left) and \PaB (right)
   in the sum-of-exclusive \acp analysis.  For $\Delta\acp$,
   similar fits are separately done for charged and neutral \PB{}'s.}
  \label{fig:acpFit}
  \vspace*{0.1in}
  \includegraphics[width=0.5\textwidth]{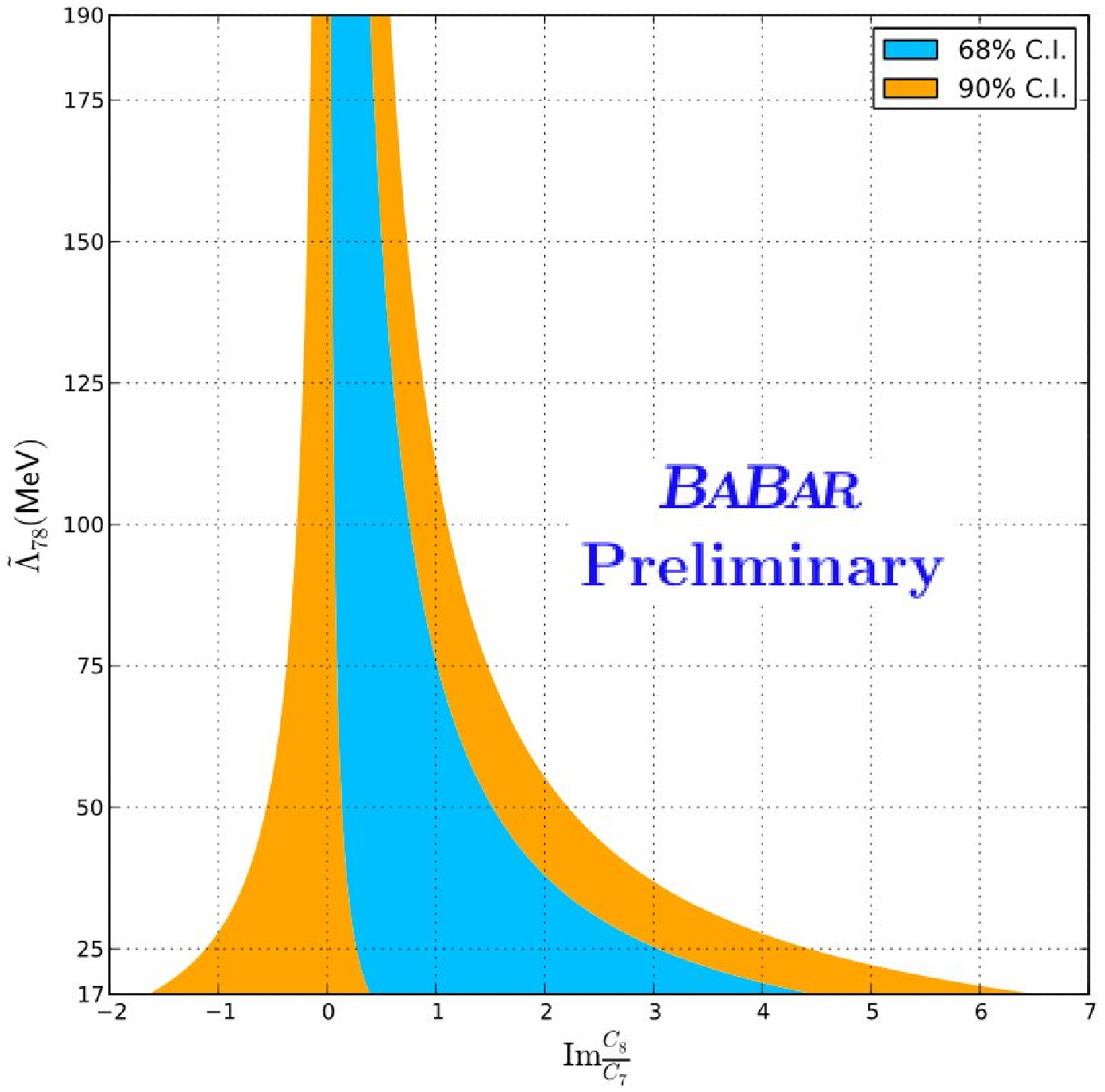}
  \caption{Regions of at least 68\% and 90\& C.L. for 
   $\Delta\acp(\PB\to\xs\Pgg)$ measurement.}
  \label{fig:c8c7}
  \vspace*{-0.2in}
 \end{center}
\end{figure}

The analysis uses standard \PB reconstruction variables,  the 
energy-substituted mass $\mes = \sqrt{E_\mathrm{beam}^{*2} - p_{\PB}^{*2}}$
and $\Delta E = E_{\PB}^* - E_\mathrm{beam}^2$, along with a set of
event-topology variables (to reduce the large continuum background)
and $\Pgg\Pgg$ mass combinations (to reduce \Pgpz backgrounds).
The selected sample represents \eg (computed most precisely from
the mass \mxs) above about 2.2\gev.
After event and best-candidate selection, the signal yield and \acp
are extracted by fits to \mes spectra for the two \CP charge states.
An example is shown in Fig.~\ref{fig:acpFit}.  The peak includes not
only signal but also signal cross-feed and a fraction of the (small)
\BB background; a systematic error on \acp of 0.009 is assigned for
possible asymmetry in the peaking background.  After correcting for
detection asymmetry, we find
\begin{equation}
  \acp(\PB\to\xs\Pgg) = 0.017 \pm 0.019\,(\mathrm{stat}) \pm 0.010\,
  (\mathrm{syst}) \ \ (\mathrm{\babar\ Preliminary})
  \label{eq:xsACPmeas}
\end{equation}
\begin{equation}
  \Delta\acp(\PB\to\xs\Pgg) = 0.050 \pm 0.039\,(\mathrm{stat}) \pm 0.015\,
  (\mathrm{syst})\ \ (\mathrm{\babar\ Preliminary})
  \label{eq:deltaACPmeas}
\end{equation}
The latter is a first measurement.  Both results are consistent with the SM.

By comparing Eqs.~\ref{eq:DeltaACP} and~\ref{eq:deltaACPmeas}, limits can
be set on $\mathrm{Im}(C_8/C_7)$, which if non-zero would be a sign of NP.
Assuming Gaussian errors, a confidence level is assigned for each value
of $\tilde{\Lambda}_{78}$ (over its full range of uncertainty) and
$\mathrm{Im}(C_8/C_7)$.  The regions of at least 68\% and 90\% C.L.
are shown in Fig.~\ref{fig:c8c7}.  The horizontal extremes of these
regions are used to set these conservative limits:
\begin{equation}
    \ 0.07 \le \mathrm{Im}(C_8/C_7) \le 4.48\ \mathrm{(68\%\, C.L.,\ 
    \babar\ Preliminary)} 
\end{equation}
\begin{equation}
    -1.64 \le \mathrm{Im}(C_8/C_7) \le 6.52\ \mathrm{(90\%\, C.L.,\ 
    \babar\ Preliminary)}
\end{equation}

\section{\boldmath{$\PB\to\Pgp\ellell$} and \boldmath{$\PB\to\Pgh\ellell$}
 Decays}

Most measurements to date of $\PB\to\xsd\ellell$ processes have been
for exclusive final states -- less precisely predicted but more easily
measured than their inclusive counterparts.  Branching fractions are
$\mathcal{O}(\alpha)$ smaller than those for $\PB\to\xsd\Pgg$.  
\xd decays are suppressed by an additional CKM factor of $\approx 23$ 
compared to \xs decays.  \babar\ has recently reported 
searches~\cite{Lees:2013lvs} for $\PBpm\to(\Pgppm/\Pgpz/\Pgh)\ellell$ 
(\Pgh to $\Pgg\Pgg$ or $\Pgpp\Pgpm\Pgpz$) from 428\invfb of data.  
SM BF predictions are in ranges
$(1.4 - 3.3)\times 10^{-8}$ for \Pgp\ellell modes, and
$(2.5 - 3.7)\times 10^{-8}$ for \Pgh\ellell. 

\begin{figure}[!t]
 \begin{center}
   \hspace{0.25in} \Pgpp\Pep\Pem \hspace{0.65in} \PKp\Pep\Pem
   \hspace{0.65in}  \Pgpz\Pep\Pem \\
  \includegraphics[width=0.7\textwidth]{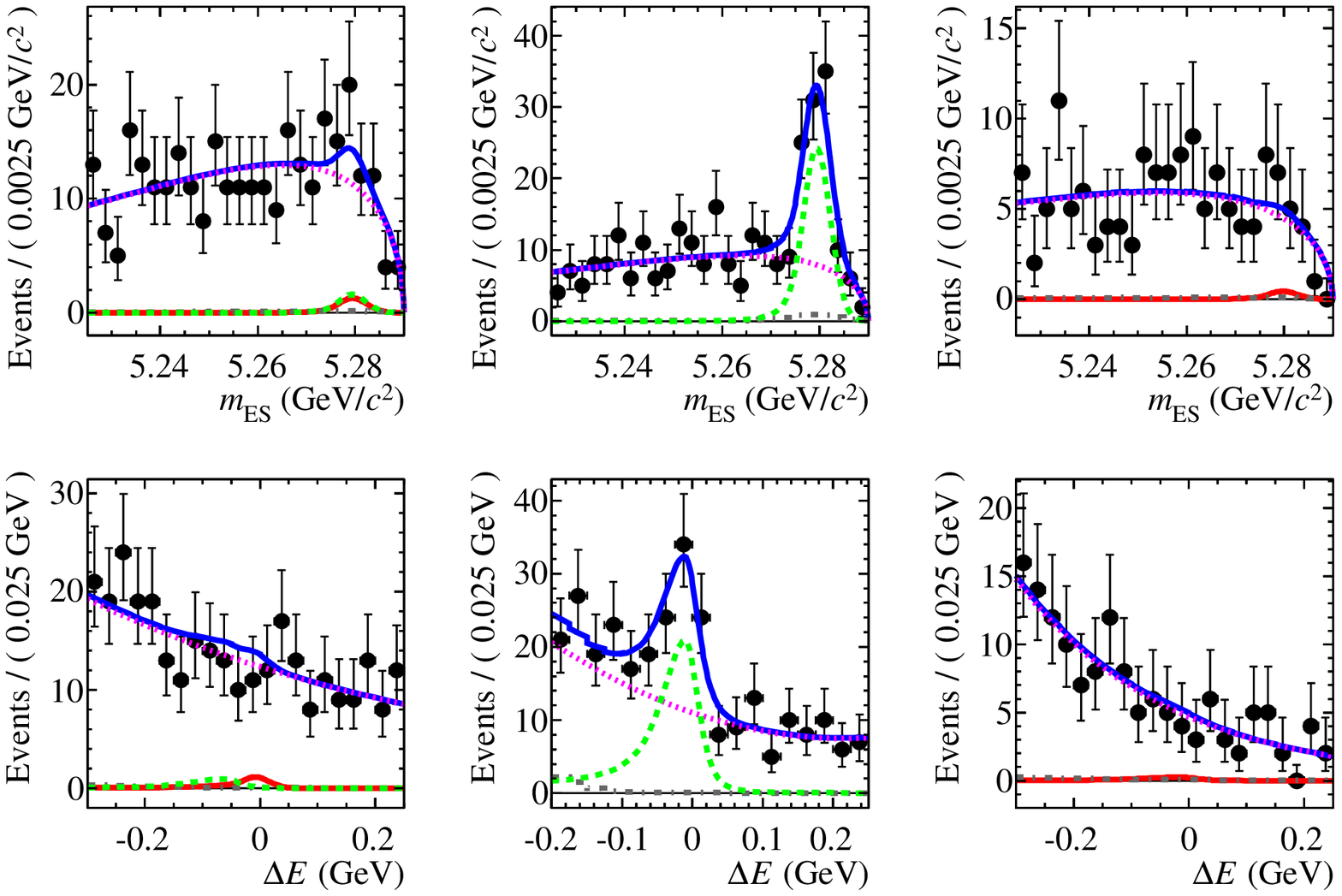}
  \vspace*{-0.1in}
  \caption{Examples of fits to data (points) for 
   \Pgp\ellell analysis.  Curves:  dotted magenta for combinatoric
   background, dot-dash gray for \PKst and \PKzS background, dashed
   green for \PKpm\ellell signal or background, and solid red for
   \Pgp\ellell signal.} 
  \label{fig:dllFits}
  \includegraphics[width=0.6\textwidth,clip]{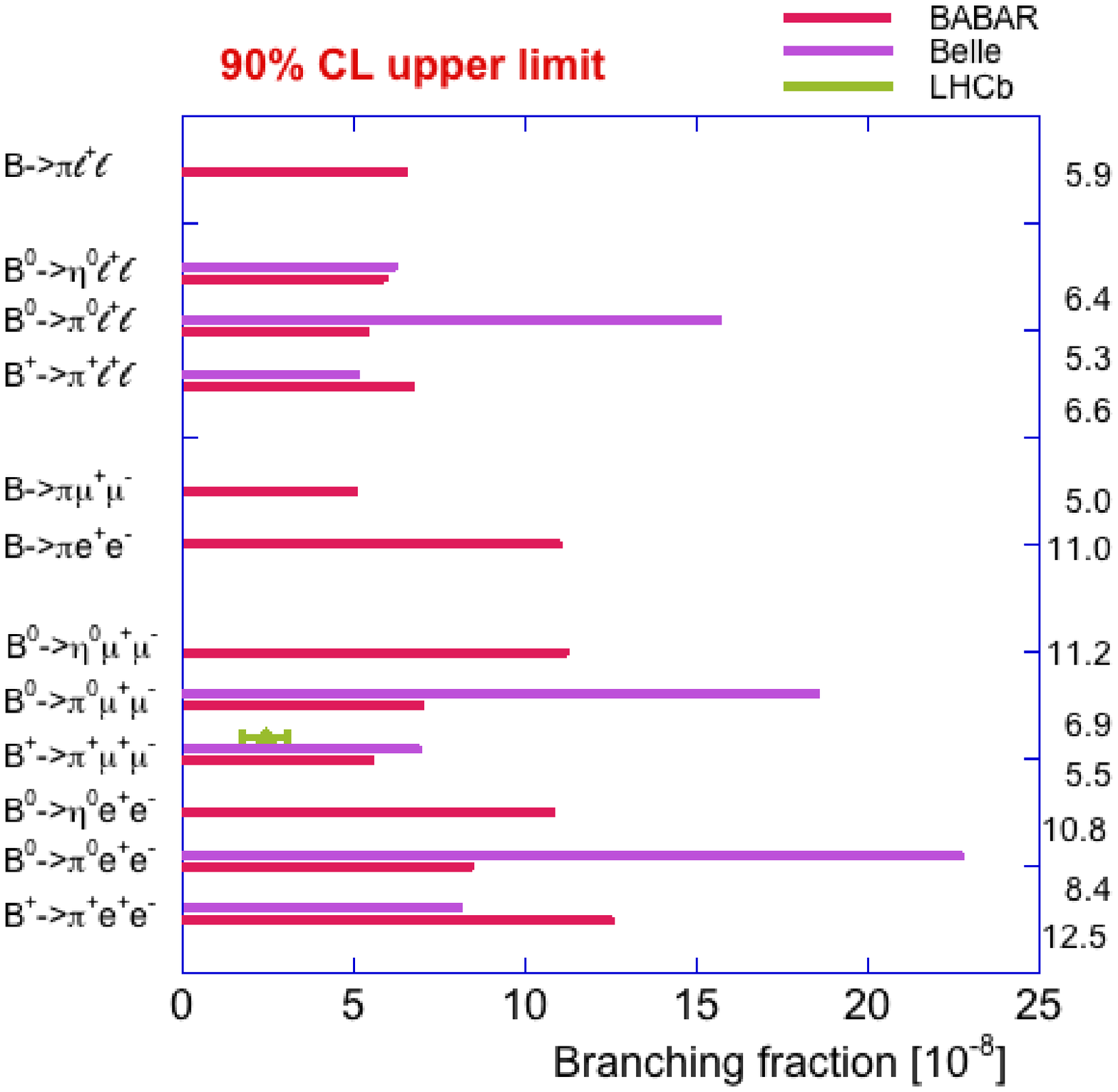}
  \vspace*{-0.2in}
  \caption{Results for $\BR(\PB\to\Pgpz(\Pgh)\ellell)$ to date.
   The \babar\ and Belle results are 90\% C.L. upper limits, with
   the \babar\ numberical values shown at the right.  The LHCB
   result is a measured value with significance $5.2\sigma$.}
  \label{fig:dllLimits}
  \vspace*{-0.2in}
 \end{center}
\end{figure}

The largest backgrounds are from:  $\PB\to\PJgy(\to\ellell) X$ 
(vetoed using $m_{\ellell}$), random combinatorics (suppressed
using event topology and missing four-momentum), and $\PB\to\kll$
with, \forex, \PKpm to \Pgppm misidentification or a lost
\Pgp from \PKzS decay (which have different $\Delta E$ spectra than
signal events).  The signal is exctracted by an unbinned maximum
likelihood fit to \mes and $\Delta E$; the decay $\PB\to\PKpm\ellell$
is also fit as signal, to constrain the misidentification background.
Some examples of the fits are shown in Fig.~\ref{fig:dllFits}.  Fits are
also done for \mumu, both lepton-flavor and pion-isospin averages, and
of course the \Pgh\ellell decays.

No signal was found in any channel, so 90\% C.L. upper limits
were determined.  They are typically about a factor of two above the
SM predicted ranges.  Figure~\ref{fig:dllLimits} summarizes these
and other measurements.  There is one positive result:  the LHCB
measurement of $\PB\to\Pgpp\mumu$ found a signal with a BF in
agreement with the SM..

\section{Summary}

Several recent \babar\ measurements had the potential for finding 
new physics (NP) beyond the SM:  $\BR(\PB\to\xs\Pgg)$;
$\acp(\PB\to\xsplusd\Pgg)$; $\acp(\PB\to\xs\Pgg)$ and $\Delta\acp(\xs)$;
a search for ultra-rare decays $\PB\to(\Pgp,\Pgh)\ellell$.
No evidence for NP was found.  The results can be used to constrain
specific NP models.  These measurements can be fruitfully
pursued at a future high-intensity \PB-factory (Belle-II).  Their power 
depends on the precision of the SM predictions and
(especially for $\BR(\PB\to\xs\Pgg)$) on
the ability to reduce systematic uncertainties.

\Acknowledgments

Results presented here would not have been possible without the
contributions of our PEP-II colleagues in achieving excellent luminosity and
machine conditions, or the expertise and dedication of the computing
organizations that have supported \babar.  I also thank the 
the SLAC National Accelerator Laboratory for its hospitality.

\end{document}

%% file: symbols.tex
\def\BR                 {{\ensuremath{\cal B}\xspace}}

\newcommand{\forex}     {\mbox{\textsl{e.g.}}\xspace}
\newcommand{\ie}        {\mbox{\textsl{i.e.}}\xspace}

\newcommand{\vs}        {\mbox{\textsl{vs.}}\xspace}

\def\ellell     {\ensuremath{\Plp\Plm}\xspace}
\def\kll        {\ensuremath{\PK^{(*)} \ellell}\xspace}
\def\xd         {\ensuremath{X_{\Pqd}}\xspace}    
\def\xs         {\ensuremath{X_{\Pqs}}\xspace}    
\def\xsd        {\ensuremath{X_{\Pqs(\Pqd)}}\xspace}    
\def\xsplusd    {\ensuremath{X_{\Pqs+\Pqd}}\xspace}    
\def\xad        {\ensuremath{X_{\Paqd}}\xspace}    
\def\xas        {\ensuremath{X_{\Paqs}}\xspace}    
\def\xasd       {\ensuremath{X_{\Paqs(\Paqd)}}\xspace}

\def\BB         {\ensuremath{\PB{}\PaB}\xspace}

\def\epem       {\ensuremath{\Pep\Pem}\xspace}

\def\tautau     {\ensuremath{\Pgtp\Pgtm}\xspace}

\def\mumu       {\ensuremath{\Pgmp\Pgmm}\xspace}

\input Upsilon.tex

 \def\eg         {\ensuremath{E_{\Pgg} }\xspace}

 \def\egcms      {\ensuremath{E^{*}_{\Pgg}}\xspace}

 \def\mes        {\ensuremath {m_\mathrm{ES}}\xspace}

 \def\mxs        {\ensuremath{m_{X_{\Pqs}} }\xspace}

 \def\bxsg       {\ensuremath{\PB \to X_{s} \Pgg}\xspace}
 \def\bxdg       {\ensuremath{\PB \to X_{d} \Pgg}\xspace}

\def\aveDelta#1 {\ensuremath{\langle \Delta_{total}#1 \rangle}\xspace}

\def\valerr#1#2#3 {\ensuremath{{#1}^{+#2}_{-#3}}\xspace}

\usepackage{relsize}
\def\babar{\mbox{\slshape B\kern-0.1em{\smaller A}\kern-0.1em
    B\kern-0.1em{\smaller A\kern-0.2em R}}}

\newcommand{\tev}{\ensuremath{\mathrm{\,Te\kern -0.1em V}}\xspace}
\newcommand{\gev}{\ensuremath{\mathrm{\,Ge\kern -0.1em V}}\xspace}
\newcommand{\mev}{\ensuremath{\mathrm{\,Me\kern -0.1em V}}\xspace}
\newcommand{\kev}{\ensuremath{\mathrm{\,ke\kern -0.1em V}}\xspace}
\newcommand{\ev}{\ensuremath{\mathrm{\,e\kern -0.1em V}}\xspace}
\newcommand{\gevc}{\ensuremath{{\mathrm{\,Ge\kern -0.1em V\!/}c}}\xspace}
\newcommand{\mevc}{\ensuremath{{\mathrm{\,Me\kern -0.1em V\!/}c}}\xspace}
\newcommand{\gevcc}{\ensuremath{{\mathrm{\,Ge\kern -0.1em V\!/}c^2}}\xspace}
\newcommand{\mevcc}{\ensuremath{{\mathrm{\,Me\kern -0.1em V\!/}c^2}}\xspace}
\newcommand{\gevbf}{\ensuremath{\mathrm{\bf \,Ge\kern -0.1em V}}\xspace}
\newcommand{\gevcbf}{\ensuremath{\mathrm{\bf \,Ge\kern -0.1em V\!/}c}\xspace}
\newcommand{\gevccbf}{\ensuremath{\mathrm{\bf \,Ge\kern -0.1em V\!/}c^2}\xspace}
\newcommand{\mevbf}{\ensuremath{\mathrm{\bf \,Me\kern -0.1em V}}\xspace}

\def\invfb   {\ensuremath{\mbox{\,fb}^{-1}}\xspace}

\def\mus  {\ensuremath{\rm \,\mus}\xspace}

\def\mus        {\ensuremath{\,\mu{\rm s}}\xspace}    

\def\to                 {\ensuremath{\rightarrow}\xspace}

\def\pep2{PEP-II}

\def\gsim{{~\raise.15em\hbox{$>$}\kern-.85em
          \lower.35em\hbox{$\sim$}~}\xspace}
\def\lsim{{~\raise.15em\hbox{$<$}\kern-.85em
          \lower.35em\hbox{$\sim$}~}\xspace}

\def\CP                {\ensuremath{C\!P}\xspace}

\def\acp        {\ensuremath{A_{\CP}}\xspace}

\def\Vtd  {\ensuremath{|V_{td}|}\xspace}

\def\Vts  {\ensuremath{|V_{ts}|}\xspace}

%% file: Upsilon.tex
\mathchardef\Upsilon="7107
\def\Y#1S{\ensuremath{\Upsilon{(#1S)}}\xspace}

%% file: CPmodes.tex
\begin{table}[!h]
 \begin{center}
  \addtolength{\extrarowheight}{1pt}
  \begin{tabular}{|l|l|} \hline
   {\color{blue} Charged Modes} & {\color{blue}\bf Neutral Modes} \\ \hline
     \PKzS\Pgpp\Pgg               & \PKp\Pgpm\Pgg           \\
     \PKp\Pgpz\Pgg                & \PKp\Pgpm\Pgpz\Pgg      \\
     \PKp\Pgpp\Pgpm\Pgg           & \PKp\Pgpp\Pgpm\Pgpm\Pgg \\
     \PKzS\Pgpp\Pgpz\Pgg          & \PKp\Pgpm\Pgpz\Pgpz\Pgg \\
     \PKp\Pgpz\Pgpz\Pgg           & \PKp\Pgh\Pgpm\Pgg       \\
     \PKzS\Pgpp\Pgpm\Pgpp\Pgg     & \PKp\PKm\PKp\Pgpm\Pgg   \\
     \PKp\Pgpp\Pgpm\Pgpz\Pgg      & \\
     \PKzS\Pgpp\Pgpz\Pgpz\Pgg     & \\
     \PKp\Pgh\Pgg                 & \\
     \PKp\PKm\PKp\Pgg             & \\
   \hline
  \end{tabular}
  \caption{The 16 modes used by \babar\ to measure $\acp(\PB\to\xs\Pgg)$.}
  \label{tab:CPmodes}
  \vspace*{-0.2in}
 \end{center}
\end{table}